\documentclass[aps,prl,reprint,superscriptaddress,showpacs]{revtex4-1}
\usepackage{graphicx}  
\usepackage{epsfig} 
\usepackage{subfigure}
\usepackage{lineno}  
\usepackage{cleveref} 

\begin{document}

\title{Measurement of the Atmospheric $\nu_e$ Spectrum with IceCube}

\affiliation{III. Physikalisches Institut, RWTH Aachen University, D-52056 Aachen, Germany}
\affiliation{School of Chemistry \& Physics, University of Adelaide, Adelaide SA, 5005 Australia}
\affiliation{Dept.~of Physics and Astronomy, University of Alaska Anchorage, 3211 Providence Dr., Anchorage, AK 99508, USA}
\affiliation{CTSPS, Clark-Atlanta University, Atlanta, GA 30314, USA}
\affiliation{School of Physics and Center for Relativistic Astrophysics, Georgia Institute of Technology, Atlanta, GA 30332, USA}
\affiliation{Dept.~of Physics, Southern University, Baton Rouge, LA 70813, USA}
\affiliation{Dept.~of Physics, University of California, Berkeley, CA 94720, USA}
\affiliation{Lawrence Berkeley National Laboratory, Berkeley, CA 94720, USA}
\affiliation{Institut f\"ur Physik, Humboldt-Universit\"at zu Berlin, D-12489 Berlin, Germany}
\affiliation{Fakult\"at f\"ur Physik \& Astronomie, Ruhr-Universit\"at Bochum, D-44780 Bochum, Germany}
\affiliation{Physikalisches Institut, Universit\"at Bonn, Nussallee 12, D-53115 Bonn, Germany}
\affiliation{Universit\'e Libre de Bruxelles, Science Faculty CP230, B-1050 Brussels, Belgium}
\affiliation{Vrije Universiteit Brussel, Dienst ELEM, B-1050 Brussels, Belgium}
\affiliation{Dept.~of Physics, Chiba University, Chiba 263-8522, Japan}
\affiliation{Dept.~of Physics and Astronomy, University of Canterbury, Private Bag 4800, Christchurch, New Zealand}
\affiliation{Dept.~of Physics, University of Maryland, College Park, MD 20742, USA}
\affiliation{Dept.~of Physics and Center for Cosmology and Astro-Particle Physics, Ohio State University, Columbus, OH 43210, USA}
\affiliation{Dept.~of Astronomy, Ohio State University, Columbus, OH 43210, USA}
\affiliation{Niels Bohr Institute, University of Copenhagen, DK-2100 Copenhagen, Denmark}
\affiliation{Dept.~of Physics, TU Dortmund University, D-44221 Dortmund, Germany}
\affiliation{Dept.~of Physics and Astronomy, Michigan State University, East Lansing, MI 48824, USA}
\affiliation{Dept.~of Physics, University of Alberta, Edmonton, Alberta, Canada T6G 2E1}
\affiliation{Erlangen Centre for Astroparticle Physics, Friedrich-Alexander-Universit\"at Erlangen-N\"urnberg, D-91058 Erlangen, Germany}
\affiliation{D\'epartement de physique nucl\'eaire et corpusculaire, Universit\'e de Gen\`eve, CH-1211 Gen\`eve, Switzerland}
\affiliation{Dept.~of Physics and Astronomy, University of Gent, B-9000 Gent, Belgium}
\affiliation{Dept.~of Physics and Astronomy, University of California, Irvine, CA 92697, USA}
\affiliation{Dept.~of Physics and Astronomy, University of Kansas, Lawrence, KS 66045, USA}
\affiliation{Dept.~of Astronomy, University of Wisconsin, Madison, WI 53706, USA}
\affiliation{Dept.~of Physics and Wisconsin IceCube Particle Astrophysics Center, University of Wisconsin, Madison, WI 53706, USA}
\affiliation{Institute of Physics, University of Mainz, Staudinger Weg 7, D-55099 Mainz, Germany}
\affiliation{Universit\'e de Mons, 7000 Mons, Belgium}
\affiliation{Technische Universit\"at M\"unchen, D-85748 Garching, Germany}
\affiliation{Bartol Research Institute and Dept.~of Physics and Astronomy, University of Delaware, Newark, DE 19716, USA}
\affiliation{Department of Physics, Yale University, New Haven, CT 06520, USA}
\affiliation{Dept.~of Physics, University of Oxford, 1 Keble Road, Oxford OX1 3NP, UK}
\affiliation{Dept.~of Physics, Drexel University, 3141 Chestnut Street, Philadelphia, PA 19104, USA}
\affiliation{Physics Department, South Dakota School of Mines and Technology, Rapid City, SD 57701, USA}
\affiliation{Dept.~of Physics, University of Wisconsin, River Falls, WI 54022, USA}
\affiliation{Oskar Klein Centre and Dept.~of Physics, Stockholm University, SE-10691 Stockholm, Sweden}
\affiliation{Dept.~of Physics and Astronomy, Stony Brook University, Stony Brook, NY 11794-3800, USA}
\affiliation{Dept.~of Physics, Sungkyunkwan University, Suwon 440-746, Korea}
\affiliation{Dept.~of Physics, University of Toronto, Toronto, Ontario, Canada, M5S 1A7}
\affiliation{Dept.~of Physics and Astronomy, University of Alabama, Tuscaloosa, AL 35487, USA}
\affiliation{Dept.~of Astronomy and Astrophysics, Pennsylvania State University, University Park, PA 16802, USA}
\affiliation{Dept.~of Physics, Pennsylvania State University, University Park, PA 16802, USA}
\affiliation{Dept.~of Physics and Astronomy, Uppsala University, Box 516, S-75120 Uppsala, Sweden}
\affiliation{Dept.~of Physics, University of Wuppertal, D-42119 Wuppertal, Germany}
\affiliation{DESY, D-15735 Zeuthen, Germany}

\author{M.~G.~Aartsen}
\affiliation{School of Chemistry \& Physics, University of Adelaide, Adelaide SA, 5005 Australia}
\author{M.~Ackermann}
\affiliation{DESY, D-15735 Zeuthen, Germany}
\author{J.~Adams}
\affiliation{Dept.~of Physics and Astronomy, University of Canterbury, Private Bag 4800, Christchurch, New Zealand}
\author{J.~A.~Aguilar}
\affiliation{Universit\'e Libre de Bruxelles, Science Faculty CP230, B-1050 Brussels, Belgium}
\author{M.~Ahlers}
\affiliation{Dept.~of Physics and Wisconsin IceCube Particle Astrophysics Center, University of Wisconsin, Madison, WI 53706, USA}
\author{M.~Ahrens}
\affiliation{Oskar Klein Centre and Dept.~of Physics, Stockholm University, SE-10691 Stockholm, Sweden}
\author{D.~Altmann}
\affiliation{Erlangen Centre for Astroparticle Physics, Friedrich-Alexander-Universit\"at Erlangen-N\"urnberg, D-91058 Erlangen, Germany}
\author{T.~Anderson}
\affiliation{Dept.~of Physics, Pennsylvania State University, University Park, PA 16802, USA}
\author{M.~Archinger}
\affiliation{Institute of Physics, University of Mainz, Staudinger Weg 7, D-55099 Mainz, Germany}
\author{C.~Arguelles}
\affiliation{Dept.~of Physics and Wisconsin IceCube Particle Astrophysics Center, University of Wisconsin, Madison, WI 53706, USA}
% ...

\date{\today}

\collaboration{IceCube Collaboration}
\noaffiliation

\begin{abstract}
We present a measurement of the atmospheric $\nu_e$ spectrum at energies 
between 0.1~TeV and 100~TeV using data from the first year of the complete IceCube detector. 
Atmospheric $\nu_e$ originate mainly from the decays of kaons produced 
in cosmic-ray air showers. This analysis selects 1078 fully contained events 
in 332 days of livetime, then identifies those consistent with particle showers. 
A likelihood analysis with improved event selection extends 
our previous measurement of the conventional $\nu_e$ fluxes to higher energies.
The data constrain the conventional $\nu_e$ flux to be $1.3^{+0.4}_{-0.3}$ 
times a baseline prediction from a Honda's calculation, 
including the knee of the cosmic-ray spectrum. 
A fit to the kaon contribution ($\xi$) to the neutrino flux 
finds a kaon component that is $\xi =1.3^{+0.5}_{-0.4}$ times the baseline value.
The fitted/measured prompt neutrino flux from charmed hadron decays strongly 
depends on the assumed astrophysical flux and shape.   
If the astrophysical component follows a power law, 
the result for the prompt flux is $0.0^{+3.0}_{-0.0}$ times a calculated flux based on the work by Enberg, Reno and Sarcevic.
\end{abstract}


\maketitle
\section{I. Introduction}
A measurement of the atmospheric neutrino flux is 
valuable in the field of neutrino astronomy and neutrino oscillation physics.
Atmospheric muon and electron neutrinos are the decay products of mesons and muons which are produced 
when cosmic-ray primaries interact in the atmosphere. 
Experiments have measured the atmospheric neutrino 
fluxes~\cite{Aartsen:2013vca,Aartsen:2012uu,Aartsen:2014qna,Diffuse-Sean-2011,Frejus-Daum-1995,SK-nu2014,SKflux-GonzalezGarcia-2006,Diffuse-Warren-2011,AMANDA-Kelley-2009,AMANDA-Julia-2010}, 
and multiple theoretical frameworks to calculate this flux are 
available~\cite{Honda-2007,Bartol-2004,Battistoni:2002ew,Charm-Enberg-2008,Martin:2003us,Naumov:1989,Sinegovskaya:2014pia}. 

Below the knee ($3\times10^{15}$~eV) of the cosmic-ray energy spectrum, the flux of $\nu_\mu$ and $\nu_e$ from $\pi$ and $K$ decays, called the {\lq conventional\rq} neutrino flux, 
follows a power law $dN/dE \propto E^{-3.7}$, where $N$ and $E$ are the number of neutrinos and the neutrino energy, respectively.
The spectral slope is steeper than that of the primary cosmic rays by about one power 
because the neutrinos' parent mesons lose a significant amount of energy in flight before decaying. 

The flux of conventional $\nu_\mu$ 
has been
measured in a wide energy range.
At  energies below several 10's of GeV, the flux is measured using fully contained events 
while, at energies above 100~GeV, flux measurements use muons produced by neutrinos traveling through the Earth, 
\textit{i.e.}~the upward-going direction.

Most $\nu_e$ come  from the semileptonic decay of charged and neutral kaons. 
The $\nu_e$ flux is lower than that of $\nu_\mu$ and
the $\nu_\mu/\nu_e$ ratio increases with increasing energy, reaching a factor of $\sim$20 at 1~TeV.
The conventional $\nu_\mu$ and $\nu_e$ flux is highest around the horizon, where parent mesons
spend a higher fraction of their lifetime at higher altitudes and are less likely to interact before they can decay.

The flux of high-energy conventional neutrinos is sensitive to the details of 
particle production in air showers.
Large uncertainties on the conventional flux models at neutrino energy above 1~TeV come from 
uncertainties in strange quark production and the cosmic ray spectrum, which
are poorly constrained by accelerator and air-shower measurements.
Precise measurements of the conventional $\nu_\mu$ and $\nu_e$ fluxes probe pion and kaon production in air showers. 

At energies between 1~TeV and 100~TeV, another class of atmospheric neutrinos 
arises, from charmed hadron decays. Since these hadrons have short 
lifetimes, the {\lq prompt\rq} neutrino flux retains the original 
spectral slope of the primary cosmic-rays.  
Prompt neutrinos are uniformly produced in the atmosphere, with equal 
fluxes of $\nu_\mu$ and $\nu_e$.  
The transition from the region dominated by the conventional 
neutrinos to the prompt neutrinos in the spectrum is expected to occur
at energies of around 1~PeV for $\nu_\mu$ and around 30~TeV for $\nu_e$.

Theoretical predictions for the atmospheric charm production have large uncertainties~\cite{Charm-Enberg-2008, Bhattacharya:2015jpa}, largely due to a lack of data on forward production at high energies. RHIC and LHC data provide useful constraints, but only in the central region. Several non-perturbative
effects come into play in the forward region of collisions. Uncertainties in the low$-x$ parton distributions and possible diffractive production channels lead to significant uncertainties~\cite{Formaggio:2013kya,Connolly:2011vc,CooperSarkar:2011pa}.

Another flux component is the astrophysical neutrinos
recently discovered by IceCube~\cite{HESE1,HESE2}.
At energies above 10's of TeVs, a seemingly isotropic flux of neutrinos 
of astrophysical origin becomes discernible 
with a spectrum harder than that of the atmospheric flux.
However, it is difficult to disentangle the prompt flux from the astrophysical component with the current event samples because their angular distributions and spectral indices are similar.
Recent IceCube analyses address the issue~\cite{Aartsen:2014muf,Gary}.

In this paper, we present a measurement of the atmospheric $\nu_e$ spectrum with IceCube.

\section{II. Detector}
IceCube is a neutrino detector deep in the South Pole ice.
The cubic-kilometer detector consists of 5,160 light sensors 
distributed on 86 vertical strings at depths between 1450~m and 2450~m below the surface. 
The array of sensors, called Digital Optical Modules (DOM)~\cite{DOM-Matis-2008}, 
observes Cherenkov light produced when charged particles exceed the speed of light in the ice. 
The DOMs consist of a pressurized glass sphere, a 252~mm diameter 
PhotoMultiplier Tube (PMT)~\cite{PMT-Mase-2010} and digitizing electronics.
Twelve LEDs in each DOM are used to calibrate the 
detector responses.

The calibration of the DOM response and the understanding of the optical properties
of the surrounding ice are crucial for the event reconstruction in IceCube.
Using \textit{in situ} LED data, the ice is modelled 
as a set of scattering and absorption parameters 
as functions of wavelength and depth~\cite{Spice-Dima-2012,AHA-Kurt-2006}. 
The ice exhibits an optically layered structure depending on dust concentration, reflecting 
 the long-term differences in climate that
affected dust accumulation over time.

The IceCube neutrino observatory includes three components, 
each designed for a specific purpose. 
The baseline array contains 4,680 DOMs on 78 strings 
with roughly 125~m string-to-string distance 
and 17~m DOM-to-DOM spacing and is optimized for detecting neutrinos 
above a few 100~GeV. 
The ``DeepCore'' sub-array  is a more densely instrumented
set of DOMs optimized for identifying neutrino events 
with energies as low as 10~GeV~\cite{JuanPablo3yrOsc}.
It contains 480 DOMs on 8 strings deployed 
in the bottom-center part of the baseline array
together with DOMs of the baseline array in the same region. 
Air showers are observed 
by a surface array called IceTop~\cite{Aartsen:2013wda}. 

The DOMs digitize the recorded PMT waveforms 
and generate time-stamped signals, or ``hits'' when the signal rises above
a threshold which is set to 0.25 photoelectrons.  
The ATWD (Analog Transient Waveform Digitizer~\cite{atwd}) 
and fADC (fast Analog to Digital Converter) digitize the waveforms 
at rate of 300 and 25 mega-samples/s,  respectively. 
The ATWD records 128 samples (430~ns total)
with a charge resolution of $\sim$30\% for single photoelectrons
and a timing resolution of $\sim$2~ns. 
The fADC system records 256 samples/event (6400~ns), to capture long, late pulses.
If a nearest or next-to-nearest neighbor DOM is also hit within $\pm$1,000~ns, then the
DOM transmits the full waveforms to the surface.  Otherwise, for isolated hits, it sends a brief summary. The isolated hits are important for efficiently recognizing incident cosmic-ray muons which might give a faint light through minimum ionization.

The surface electronics forms a trigger when  
at least eight non-isolated hits are observed in a 5,000~ns window. Then, a physics event is built, containing all of the isolated and non-isolated hits.  
Further details about the detector can be found in 
Ref.~\cite{Achterberg:2006md,Halzen:2010yj,Collaboration:2011ym}.

The IceCube coordinate system is right-handed, with its origin at the center of the baseline array,
with the $z$-axis pointing upward. The $y$-axis follows the prime (Greenwich) meridian, and the $x$-axis
points toward $+$90 degrees longitude. The zenith angle ($\theta$) is defined in the usual manner,
the angle between the event arrival direction and the $z$-axis, while the azimuthal angle ($\phi$)
is measured from the positive $x$-axis, in the $x-y$ plane.

\section{III. Data and Simulation}
This analysis uses data taken with the full 86-string configuration of IceCube,
between May 13, 2011 and May 15, 2012. After excluding calibration runs and a few periods when the detector was operating in a partial configuration or exhibiting large variations in rate, the 
livetime is 332.3 days. This selection avoids systematic biases 
due to detector instability and ensures all strings of the detector are active. 

In order to avoid statistical bias, the analysis cuts and fit procedure were developed
using only 10\% of the data, spread evenly throughout the year.  After the cuts and fit were fixed, the rest of the data was studied.

For this analysis, the signal is defined as atmospheric $\nu_e$ 
interactions contained inside the detector volume.
Containment criteria are based on the vertex position, which is determined using both
the first DOM hit in time and a vertex reconstruction.
Non-contained background events entering from 
outside the detector are vetoed by these containment criteria (Section IV). 

When a high-energy neutrino interacts in the ice,
the deep inelastic scattering can produce one of three event signatures.
{\lq Cascades\rq} are created by $\nu_e$ charged current~(CC) 
interactions which consist of an electromagnetic shower 
and a hadronic shower, or neutral current~(NC) interactions of all neutrino flavors.
{\lq Tracks\rq}  are through-going muons 
from $\nu_\mu$~CC interactions occurring outside of the detector.
{\lq Hybrid\rq}  events from $\nu_\mu$~CC interactions occurring
within the detector have both a hadronic shower and a track.

The Cherenkov light yield of the shower particles is proportional to the cascade energy.
Hadronic showers have lower light output and larger shower-to-shower variations than electromagnetic showers~\cite{Radel:2012ij,Kowalski-Thesis-2004}. This is partly because hadrons are heavier than electrons, with higher Cherenkov thresholds.
Also, hadronic showers produce neutral particles, have nuclear interactions, and transfer energy to struck nucleons.
A 1~TeV hadronic shower has a light output which is $(80\pm10)\%$~\footnote{
  $F$ is a ratio of the light yield of the hadronic to electromagnetic cascades of the same energy $E$ in GeV.
  $F = 1-(E/E_0)^{-m}\cdot(1-f_0)$,
  where $E_0$ = 0.40, $m$ = 0.13, and $f_0$ = 0.47.
  The uncertainty is $\sigma = F \cdot \delta_0 \cdot (\log_{10}E)^{−-\gamma}$,
  where $\delta_0$ = 0.38 and $\gamma$ = 1.16~\cite{Spice-Dima-2012}}
of that of an electromagnetic shower of the same energy~\cite{Kowalski-Thesis-2004}.  In the simulations,
a parameterization is used to account for the reduced light output.   The visible energy ($E_{vis}$) is defined as the observed energy, assuming that the shower is electromagnetic.  
The pattern of detected light is roughly spherical for both types of showers due to short travel lengths of the shower particles.

The largest background in this analysis is downward going muons 
produced by high-energy cosmic ray interactions ({\lq CR muons{\rq).
The CR muons which reach the surface with an energy of 500~GeV or larger can 
penetrate the ice to the depth of IceCube, and can become a background to 
the atmospheric neutrino signal.
This muon background has three main signatures.
The first kind is from through-going tracks created outside 
of the detector.  This could be a down-going single muon or muon bundle 
from a cosmic-ray interaction. 
The second kind is an event with multiple tracks 
having different directions produced by coincident 
but unrelated air showers. 
The third kind is a {\lq stealth{\rq~muon which 
passes between strings or through the dustiest, optically most absorbing ice layers. 
This class of events has the appearance of the cascade signal
when the muon generates only a few hits 
in the outer regions of the detector and then 
undergoes stochastic losses that release 
most of its energy in a cascade-like shower within the fiducial region. 
There is also a small background from through-going muons from $\nu_\mu$ interactions outside of the detector.

Air showers are simulated with CORSIKA (COsmic Ray SImulations for KAscade)~\cite{CORSIKA-Heck-1998} including the Sibyll~\cite{Ahn:2009wx} hadronic interaction model. 
In IceCube the cosmic ray spectra are simulated for five nuclei. 
By re-weighting the five spectra, 
a resulting muon flux is obtained to represent a cosmic ray composition model.
In this analysis, we used the phenomenological {\lq H3a\rq} 
composition model~\cite{Gaisser:2012zz} which takes into 
account updated cosmic ray spectra
and the most recent spectral slope measurements~\cite{cream,Ahn:2010gv,Ahn:2009gv}.
IceCube data are in good agreement with the H3a model predictions
in the energy range relevant for this analysis (1$-$1000~TeV in the primary cosmic-ray energy). 
An alternative model, the poly-gonato spectrum~\cite{PolyGonato-Hoerandel-2003} also models cosmic rays with five different nuclei.  It uses different parameterizations, particularly for the knee, and finds, for this analysis, a roughly 30\% difference in the background normalization. 

Neutrinos are simulated by using a software module (\texttt{neutrino-generator}) which is based on 
the ANIS package~\cite{ANIS-Gazizov-2004} with CTEQ5~\cite{Lai:1999wy} 
cross-section tables. 
To obtain a large sample of simulated neutrinos, 
the simulation forces all neutrinos to interact 
near the detector and then assigns each an interaction probability. 
The conventional flux from Ref.~\cite{Honda-2007} ({\lq Honda{\rq) was 
used up to 10~TeV. 
The Honda model is extrapolated to higher energies 
using the flux parameterization~\cite{crbook} 
\begin{eqnarray}  \label{eq1}
  \Phi(E_{\nu}) &=& C \cdot E^{-\alpha}_{\nu} \cdot ( w_\pi + w_K), \\
  w_\pi &=& \frac{A_{\pi \nu}} {1+ B_{\pi \nu}E_{\nu}\cos{\theta^*}/\epsilon_{\pi}}, \\
  w_K &=& \frac{A_{K \nu}} {1+ B_{K \nu}E_{\nu}\cos{\theta^*}/\epsilon_{K}}.
\end{eqnarray}
The $w_\pi$ and $w_K$ are relative contributions to the neutrino flux from $\pi$ and $K$, respectively.
The parameters $A$, $B$, and the absolute normalization ($C$) of the flux are determined by fitting to the published Honda flux 
at lower energies and $\theta^*$ is the neutrino zenith angle at the production point.
The index and critical energies are $\alpha=2.65$, $\epsilon_{\pi}=$ 115~GeV, 
and $\epsilon_{K}=$ 850~GeV. 
For prompt neutrinos, the flux from Ref.~\cite{Charm-Enberg-2008} ({\lq ERS{\rq) is used. 
Both the conventional and prompt baseline predictions are corrected to an updated cosmic-ray spectrum, including a knee structure which is similar to the H3a spectrum used for the cosmic-ray simulations. We also apply a small correction factor of 0.5\% to account for the additional $\nu_e$ production from
$K_s$ semileptonic decays~\cite{Kshort}.
Tau neutrinos are not included as an atmospheric component.
The contribution is less than 5\% compared to the total prompt contribution
because its flux mainly comes $D_s$ decays which are smaller
than other charmed hadron decays in the atmosphere~\cite{Charm-Enberg-2008}.

With increasing energy, the probability of vetoing an atmospheric neutrino
through the presence of CR muons from the same cosmic-ray shower
increases in the downward region. 
Since the two coincident particles are nearly collinear, 
the events are automatically rejected in analyses sensitive 
to the downward contained events and therefore the veto probability as an additional correction should be applied to the event rate~\cite{Schonert:2008is,Gaisser:2014bja}.

The efficiency of this atmospheric self veto correction depends on the neutrino production processes.
There are two types of veto.
A {\lq correlated{\rq} veto occurs when a muon and a neutrino are produced in the same decay, while for a {\lq non-correlated{\rq} veto they have different parent particles from the same shower.
For the conventional $\nu_\mu$, the correlated veto is the dominant process 
while, for the conventional $\nu_e$, the non-correlated veto is the main process.
For the prompt neutrinos at energies well above the $\epsilon_{K}$, 
the impact from the non-correlated component increases 
since decay modes of the parent charmed hadrons involve fewer number of correlated muons.
The reference model~\cite{Gaisser:2014bja} used here
treats the correlated and non-correlated components for $\nu_\mu$ and $\nu_e$ separately. 
As Fig. 3 of Ref.  ~\cite{Gaisser:2014bja}  shows, for neutrinos above 10~TeV in the vertically downward region, the veto probability 
is higher than 95\% while at $\cos(\theta) = 0.2$,
the probability is close to 50\%.

In the following, the {\lq modified Honda{\rq} flux includes
the extrapolation to higher energies of the original Honda flux with the input H3a spectra
and the additional $K_s$ contribution.
Similarly, the {\lq modified ERS{\rq} flux refers to the ERS flux with the H3a spectra.
These modified fluxes are used to weight the neutrino 
simulations throughout this analysis (details can be found in~\cite{Anne}.) 
For baseline event rate predictions, the self veto corrections are applied to these 
modified fluxes.
To ensure consistency, the conventional neutrino expectations 
are partly validated against the neutrino predictions from 
the CORSIKA generator in the downward region where 
the self veto is in effect. 
The astrophysical flux is modelled as a single power law 
${dN/dE} \sim \phi_0 \cdot (E/100\rm TeV)^{-\gamma}$, where $\phi_0$ is the flux at 100~TeV and $\gamma$ is the spectral index. 

The simulated particles are propagated to the detector, and 
Cherenkov photons are produced from the 
charged particles.
The generated photons are tracked through the ice, using the 
measured optical scattering and absorption coefficients, and then through a simulation of the hardware response.
For example, at $z=-350$~m, a Cherenkov photon can travel 200~m with on average three scatters.
IceCube simulates photons from an event in a cylindrical volume which is roughly twice bigger in radius and length
than the instrumented volume to maximize the light collection efficiency for events
skimming the edge of the detector.
The Monte Carlo events use the same format as data events.  Both are treated identically in event processing.
The average trigger rate was 2200 Hz with a roughly 10\% seasonal variation due to temperature and pressure changes
in the atmospheric conditions above IceCube. 
The CR muon rate at trigger level ({\lq Level~1{\rq}) is $7.3 \times 10^6$ times 
larger than that of the atmospheric $\nu_e$ signal, 
which is predicted to be $3.0 \times 10^{-4}$ Hz above 300~GeV.

\section{IV. Event Selection}
The event selection proceeds in several stages 
to enrich the atmospheric $\nu_e$ signal against 
the large CR muon background by contrasting the simulated 
signal with background Monte Carlos. 
As it is currently not possible to distinguish electromagnetic showers 
from hadronic ones in IceCube, the sample contains background from NC interactions of other neutrino flavors. 
The selection relies on the searches for the spherical 
hit pattern of light in the cascade signal 
and reconstruction variables which describe the cascade signal in ice.
The cascade variables used in this analysis are 
explained in more details in Ref.~\cite{Aartsen:2013vca,AMANDA-Christopher-2004}.

\subsection{Reconstructions}
Two maximum-likelihood algorithms are used to reconstruct events under the cascade hypothesis.
The first estimate ({\lq Cascade-LLH {\rq)~\cite{Kowalski-Thesis-2004,Cascade-Mike-2011} 
of a cascade interaction position ($X_{\rm vertex}$, $Y_{\rm vertex}$, and $Z_{\rm vertex}$), 
and an interaction time uses only hit time information.
The algorithm uses an analytic probability density 
function (PDF), $p(t_{res},d_i)$~\cite{Pandel-Thesis-1996} expressed in inverse nanoseconds, 
for constructing a likelihood :
\begin{eqnarray} \label{eq_tres}
  \mathcal{L} = \prod_{i=1}^{\rm hits} p(t_{res},d_i) \\
  t_{res} = t_i - t_{geo} = t_i - t_0 - d_i / c_{ice},
\end{eqnarray}
where $t_i$ is the observed time of the hit, $t_0$ the expected time 
of the cascade interaction, $d_i$ the distance 
from the hit DOM to the interaction vertex, and $c_{ice}$ the speed of light in ice. 
The time delay of a hit relative 
to the geometrical time ($t_{geo}$) corresponding to straight-line propagation is defined 
as the residual time, $t_{res}$, \textit{i.e.} a non-scattered photon registers at $t_{res}=0$.
Cascade-LLH provides an initial vertex seed for an improved reconstruction 
and returns a cascade quality parameter $RLLH_{vertex} = -\log \mathcal{L}/(N_{hit} -4)$, 
an analog of reduced $\chi^2$, with a number of hits ($N_{hit}$) minus four degrees of freedom. 
The calculation assumes that photon scattering is independent of depth in the ice, so
the reconstruction has a limited resolution.
The resolution is measured as a Gaussian spread of 1~$\sigma$ 
on the difference between a true cascade vertex and a reconstructed vertex.
The Cascade-LLH reconstructs the interaction vertex 
with a resolution of 11~m in the $x$-$y$ plane and 12~m in $z$
for 10~TeV $\nu_e$.

The second, more advanced algorithm, ({\lq CREDO {\rq)~\cite{Aartsen:2013vca,Aartsen:2013vja,masterthesis:middell} 
reconstructs seven parameters of a cascade in a single fit. 
The vertex position ($X_{\rm reco}$,$Y_{\rm reco}$, and $Z_{\rm reco}$), 
the time ($t_{\rm reco}$), the direction ($\theta_{reco}$ and $\phi_{reco}$) and 
the visible energy ($E_{reco}$) of the cascade are estimated 
by using full waveform information. 
CREDO uses a more detailed PDF for time and amplitude expectations that includes the 
depth dependent propagation of light in ice. 
The scattering and absorption properties are stored 
in a table which is interpolated with 
splines~\cite{Whitehorn:2013nh}. 
The vertex resolution of CREDO is 4~m in the $x$-$y$ plane and 3~m in $z$ for 10~TeV $\nu_e$.

Likelihood reconstructions based on the track hypothesis~\cite{AMANDA-Christopher-2004} are used to identify
the CR muon background events and to estimate their direction ($\theta_{track}$ and $\phi_{track}$). 
As a muon travels close to the speed of light, 
a likelihood similar to Eq.~\ref{eq_tres} with a new $t_{geo}$ definition is fitted.
The PDF used in the reconstruction is called a single photoelectron (SPE) PDF 
which models the $t_{res}$ using only the earliest photon at each DOM.
Additionally, a zenith-weighted Bayesian track reconstruction is performed
using prior knowledge of the CR muon angular distribution.
Only the downward-going direction is allowed for the reconstructed track directions since the reconstruction maximizes the product of the PDF
and the prior.

\subsection{Level 2 (Cascade Online Filter)}
The events recorded at the South Pole are filtered to reduce the data volume 
so that the data 
can be transferred to off-sites via satellite. 
This online filter (Level~2) algorithm removes early and late PMT hits 
unrelated to physics interactions.
The remaining hits are used to calculate 
simple topology variables and $RLLH_{vertex}$ which are used in the filtering.
The background rejection factor is 99\% with the filter retaining 77\% of 
the atmospheric $\nu_e$ signal above 300~GeV. The efficiency reaches 90\% above 10~TeV.
The CR muon background after Level 2 selection comprises 
about 60\%, 20\%, and 20\% of through-going muons, 
coincident muons, and stealth muons, respectively. 

\subsection{Level 3 (Containment)}
The containment cuts require that the cascade vertex is in the fiducial region.
In addition to simple containment conditions based on the earliest hit time, cuts based on the vertex reconstruction
are applied to make the light produced by the cascades contained within the detector volume. This Level~3 filter reduces the CR muon background further as the background-to-signal ratio is still high ($\sim10^5$).

An algorithm identifies clusters of hits which are distinct in time and space. Only the events classified as a single cluster are accepted, in order to reject coincident CR muon background. 

The first hit must not be on one of the outer strings, and must be no closer than 70~m
to the top or bottom of the detector, \textit{i.e.}  $-430~\rm m<Z_1<430~\rm m$.

The fiducial volume cut requires the reconstructed vertex from CREDO ($X_{\rm reco}, Y_{\rm reco}, Z_{\rm reco}$) must be within a cylinder of 420 m radius from the center of the detector with 70~m minimum distance to the edge of the detector.
Additionally, the vertex should be no closer than 100~m to the top or 50~m to the bottom of the detector ($-450~\rm m< Z_{\rm reco}<400~\rm m$) . 

For the contained events, we further impose several quality cuts. 
Each event must have hits on at least three non-DeepCore strings. 
The ratio of the number of hit DOMs to the total number of DOMs within a sphere centered on the vertex should be high ($>$60\%)~\cite{Cascade-Joanna-2011,Aartsen:2013vca}.
The radius of the sphere is determined by the root-mean-square distance to the vertex of hit DOMs with a scaling factor
that maximizes signal selection power.
Then, events with a low $RLLH_{vertex}$ are selected.
After this Level~3 selection, the dominant CR muon background is 
stealth muons with a few veto hits. 
These muons are typically minimally ionizing in
the veto region and then produce a stochastic signature 
in the fiducial region, mimicking the cascade signal.
The cut efficiencies will be discussed later.

\subsection{Level 4 (Neutrino Selection)}
The Level~4 event selection uses a machine learning technique 
to separate the atmospheric 
cascade signal from the CR muon background. 
A multivariate analysis method based on boosted decision trees (BDT) 
is implemented using a toolkit for multivariate 
data analysis~\cite{Hocker:2007ht}. The BDT uses 12 variables, 
chosen for their power to separate cascades 
from the CR muon backgrounds.
The variables are listed below, classified in three categories: veto, quality, and topology.
\begin{description}
  \item[Veto] \hfill \\
    \textbf{1.~$N_{veto}$}: The number of hits recorded before the CREDO vertex in time and consistent with downward-going muons.\\
    \textbf{2.~$N_{cone}$}: The number of hits in a cone with its apex at the CREDO vertex position, an opening angle of 36 degrees, and centered on the incoming track direction obtained from the SPE reconstruction. \\
    \textbf{3.~$Z_{reco}$} \\
    \textbf{4.~$\rho_{reco}=\sqrt{X_{\rm reco}^2+ Y_{\rm reco}^2}$}\\

  \item[Quality] \hfill \\
    \textbf{5.~$\theta_{track}$}: The zenith angle of the SPE reconstruction.\\
    \textbf{6.~$\theta_{reco}$}: The zenith angle of the CREDO reconstruction.\\
    \textbf{7.~$RLLH_{vertex}$} \\
    \textbf{8.~$R_{likelihood}$}: The likelihood ratio of the SPE-track hypothesis to the Cascade-LLH hypothesis.\\
    \textbf{9.~$R_{Bayes}$}: The likelihood ratio of downward forced track hypothesis (zenith-weighted SPE)
    to non-forced track hypothesis (SPE).\\

  \item[Topology] \hfill \\
    \textbf{10.~$R_Q$}: The charge fraction in the first 300~ns, excluding the two earliest hits. \\
    \textbf{11.~$Z_{split}$}: The vertical distance between hit 
    center-of-gravities determined by splitting in time into two clusters of hits.\\
    \textbf{12.~$Z_{speed}$}: The $z$-coordinate component of a reconstructed velocity calculated 
    using the first half of all hits in an event.\\
\end{description}

The discriminating power comes relatively evenly from the three categories. 
The most powerful separators are $R_{likelihood}$, $Z_{split}$, and $N_{cone}$ 
which not only describe the data well but also show minimal 
correlations with other variables. 
The distributions of these variables are shown in Fig.~\ref{l3_var}. 
\begin{figure*}[!t]
  \centering
  \includegraphics[width=6.5in]{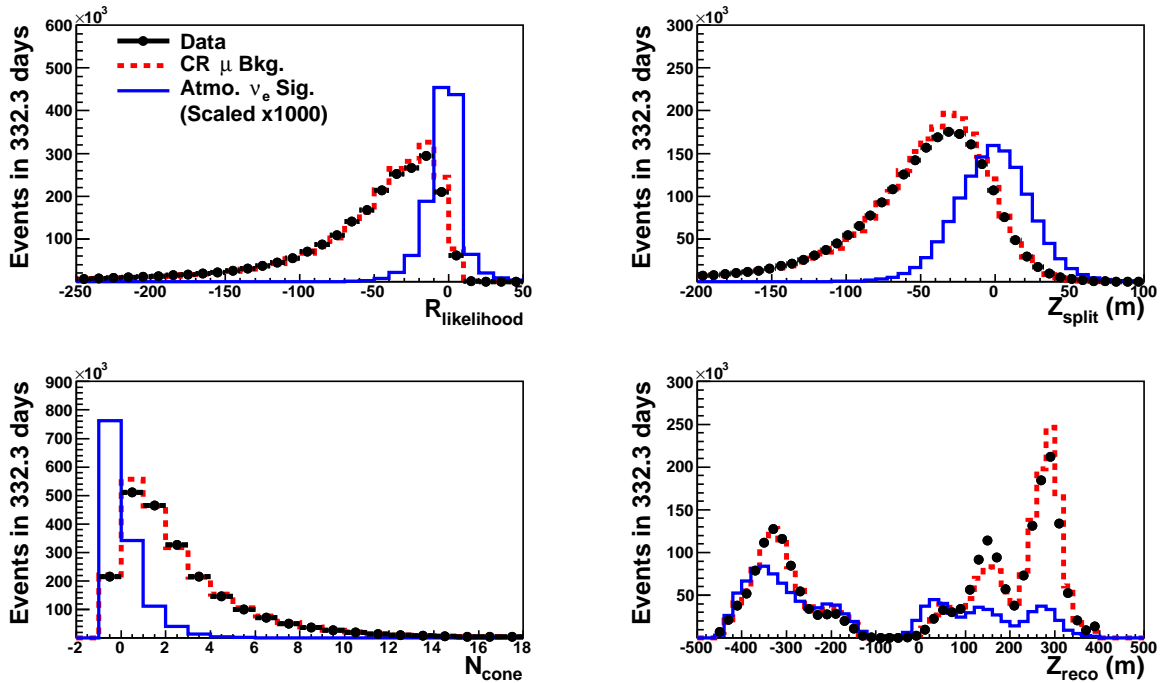}
  \caption{Four representative variables in the BDT (see description in the text). 
The dotted red lines are for CR muon background simulation 
while the blue lines are for the atmospheric $\nu_e$ simulations. 
Real data are shown with black circles. 
The $\nu_e$ signal is scaled up by 1000 to show the separation.}
  \label{l3_var}
\end{figure*}
The Veto variables ensure that the contained events show no trace of an incoming muon track before a reconstructed vertex time. The Quality variables identify whether the hit pattern is 
extended in any particular direction (track-like) or is isotropic (cascade-like).
The Topology variables look for cascade events with more localized charge distribution to distinguish them from the long through-going events that distribute hits in a larger distance.
The BDT output is a discriminant score for each event, with a higher number 
indicating more signal-like events. 

The distributions of the BDT scores
are shown in Fig.~\ref{l3bdt}. 
The data and the total Monte Carlo prediction display a transition from the region dominated by the CR muons
to the atmospheric neutrino dominant region. 
The gradual over-prediction of the CR muon background going from a low BDT score to a high score
is mainly due to the limitations in the modelling of the detector systematic uncertainties.
At high BDT scores, the CR muon background events are relatively 
more populous at the bottom part of the detector
where our veto is less effective in rejecting the CR muon simulation events than in data. 
These events come through the dustiest ice region between 2000~m and 2100~m in depth with shallow zenith angles and produce few hits 
in the veto region. Since the BDT uses depth-dependent 
variables such as $Z_{reco}$, it is sensitive to absorption of 
the photons in ice. 

The final selection is based on 
the two-dimensional BDT-Energy cut shown 
in Fig.~\ref{EBDT}.  One limitation of this selection is that it was based on CR muon
simulations which had limited statistics (about 10\% of the data live time), set by the
available processing power.
For this reason, the CR muon background was not estimated from the CORSIKA sample using the final BDT-Energy cut.
Table~\ref{table_level} summarizes the event selection. 
The efficiency is shown in Fig.~\ref{veff}.
\begin{figure}[!t]
  \centering
  \includegraphics[width=3.5in]{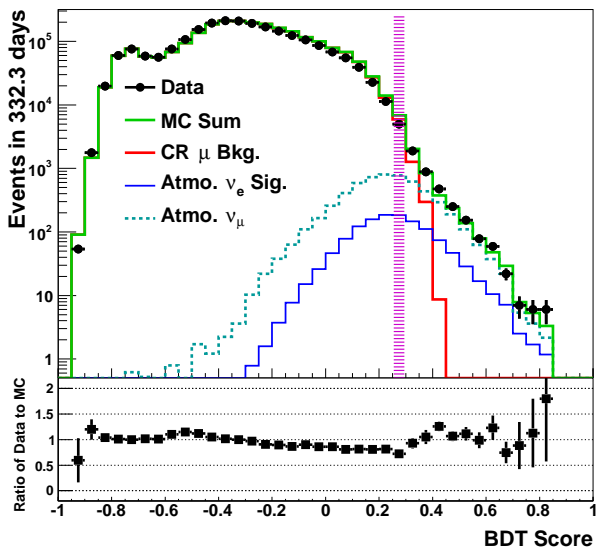}
  \caption{BDT score distribution at Level 3. Real data are shown with points 
    while the sum of all Monte Carlo prediction is shown as a solid green line. 
    Atmospheric neutrino predictions are shown with the cyan dotted line ($\nu_\mu$) 
    and the blue solid line ($\nu_e$). The CR muon background simulation is 
    shown with a red line. The side band for the final level CR muon estimation 
    is indicated with the magenta vertical band.}
  \label{l3bdt}
\end{figure}
\begin{figure*}[!t]
  \centering
  \includegraphics[width=6.5in]{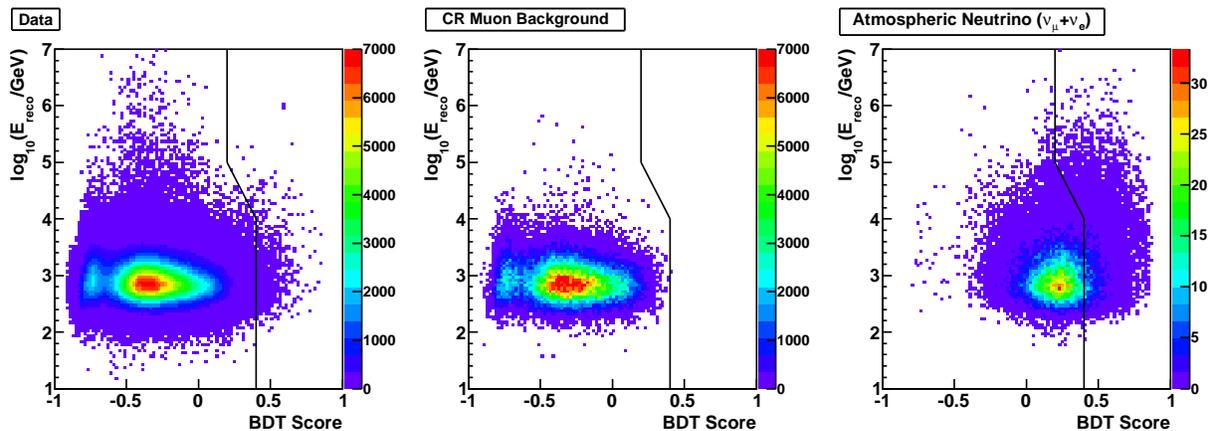}
  \caption{Two-dimensional distributions of a reconstructed 
    visible energy as a function of the BDT score are shown for the real data (left), 
    CR muon background (middle), and atmospheric neutrinos (right). 
    The right side of the black vertical line indicates the final event sample. 
    The z-axis is the number of events in 332.3 days.}
  \label{EBDT}
\end{figure*}
\begin{table}[t]
\begin{center}
\caption{The number of events in 332 days of data are shown at each level.
  The Level~4 rates are the final best fit values where CR muons are estimated from the data side band. 
  The total neutrino rates are based on the modified Honda predictions. 
  Level 1 \& 2 neutrino rates are only for energies above 300~GeV.
  $\nu_e$ contains about 8\% NC events at all levels but $\nu_\mu$ NC fraction($R_{NC}$) 
  grows as the higher level selection enriches cascades.}
\begin{tabular}{c|r|r|r|r}
\hline\hline
Level& Data & CR muon & Atm. $\nu_{e}(R_{NC}\%)$ & Atm. $\nu_{\mu}(R_{NC}\%)$ \\
\hline
1& $6.3\times10^{10}$ &  $5.2\times10^{10}$  & $8.6\times10^{3}$(10\%) & $2.3\times10^{5}$(6\%) \\
2& $7.8\times10^{8}$  & $5.8\times10^{8}$  & $6.6\times10^{3}$(8\%) & $1.0\times10^5$(14\%) \\
3& $2.3\times10^{6}$  & $2.3\times10^{6}$ & 1278(8\%) & 5784(35\%)\\
4& 1078               & 115 & 215 (8\%) & 645 (40\%) \\
\hline\hline
\end{tabular}
\label{table_level}
\end{center}
  \vspace{-5mm}
\end{table}
\begin{figure}[!t]
  \centering
  \includegraphics[width=3.in]{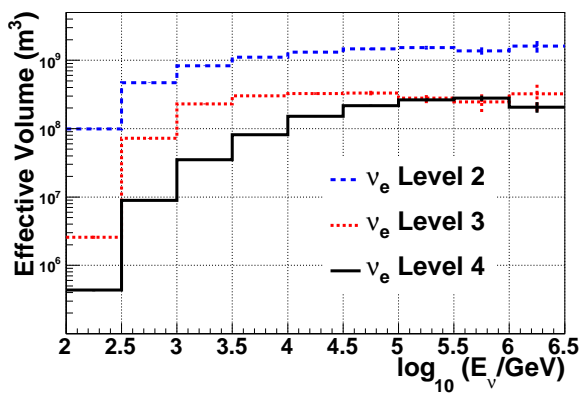}
  \caption{The $\nu_e$ effective volume is plotted as a function of energy at different cut levels. The effective volume is the detector volume multiplied by the ratio of the number of selected events to the number of generated events over 4~$\pi$ steradian.
 No self veto correction is applied to the number of events. The relative efficiency change 
 at high energies is mainly due to containment requirements 
 while, at low energies, the change is driven by the strict selections needed to reject backgrounds.}
  \label{veff}
\end{figure}

To get a more accurate estimate of the CR muon background 
without relying on simulations, we estimate the final rate
from background-dominated data close to the signal region.
As Fig.~\ref{EBDT} shows, the region around 10~TeV is 
poorly populated in the CR muon background simulation.
However, the BDT score shows no strong dependence on the reconstructed energy. 
This is expected because the training variables do not contain explicit 
energy information.
The data region for the background estimation is chosen such that sample size is maximized and neutrino contamination is minimized while staying as close as possible to the high-BDT score signal region.
We obtain the optimal control sample with BDT scores between 0.25 and 0.3, 
as shown by the vertical band in Fig.~\ref{l3bdt}. 
Observable distributions for data in this region are used as templates for the final CR muon background.
To check the robustness of this background estimate, distributions of neighboring data regions with the same width have been evaluated. Results using these alternative bands show no significant deviation from the baseline choice.

\section{V. Particle Identification}
After Level~4, particle identification (PID) variables are used
to distinguish hybrid ($\nu_\mu$~CC) events from cascades. 
For a more accurate vertex position, angular, and energy reconstruction of cascades, 
we use slower {\lq}iterative{\rq} CREDO reconstruction~\cite{Aartsen:2013vca,AMANDA-Christopher-2004,masterthesis:middell}.
The CREDO reconstruction with four different angular seeds mitigates the probability of the minimizer becoming trapped in local 
minima, a common problem in scenarios with a high number of dimensions.
The improved results for a reconstructed energy ($E_{reco, 4}$) and for the reconstructed zenith angle ($\theta_{reco, 4}$) are shown in Fig.~\ref{res} and used in the analysis fitting procedure.

For signal $\nu_e$~CC events at around 10~TeV, 
the energy resolution is $\Delta E_{vis}/E_{vis}\approx\pm9\%$, 
where $\Delta E_{vis}=E_{vis} - E_{reco, 4}$. The mean of the $\Delta E_{vis}$ distribution
overestimates the visible energy by about 6\%, but, because of neutral-current interactions,
$E_{vis}$ is on average 4\% lower than the true neutrino energy. 
For the same energy range,
the zenith angle accuracy is $\Delta \theta_{\nu}/\theta_{\nu}\approx\pm8$~degrees, 
where $\theta_{\nu}$ is the true zenith angle of the neutrino and 
$\Delta \theta_{\nu}$ is the difference between the neutrino zenith angle and $\theta_{reco, 4}$.
The mean of the $\Delta \theta_{\nu}$ distribution does not show a bias.

Systematic effects add additional uncertainties on the resolutions which are evaluated using alternative simulations.
The optical efficiency of a DOM and the optical properties of ice are varied for those simulations by a known amount.
We treat a maximum deviation from the baseline simulation as a size of the uncertainty. From this study, 12\% for energy uncertainty and 2 degrees for zenith angle uncertainty are obtained.
\begin{figure}[!t]
  \centering
  \includegraphics[width=3.5in]{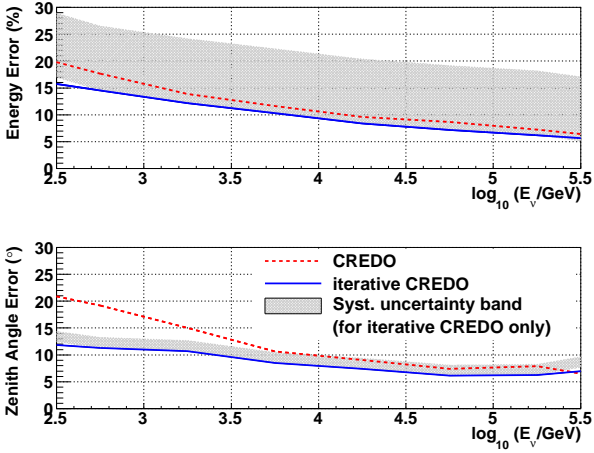}
  \caption{Cascade reconstruction uncertainties as a function 
    of true $\nu_e$ energy using $\nu_e$~CC interactions. The upper panel shows the energy statistical (only) resolution 
    obtained by comparing the visible energy with reconstructed energy. 
    The lower panel is for zenith angle resolution in degrees. 
    Solid blue lines represent the iterative reconstruction results. 
    Systematic uncertainties (gray bands) add 
    an additional 12\% for energy uncertainty and 2 degrees for zenith angle uncertainty.}
  \label{res}
\end{figure}

With the iterative CREDO results, four selected variables are 
used to train a BDT for particle identification (PID-BDT) by treating $\nu_\mu$~CC events as background. 
These variables exploit hits originating from the muon in the $\nu_\mu$~CC interaction.   Two variables depend on the first photon arrival times at the DOMs, relative to the time expected for a point-like emitter (cascade hypothesis) for a photon that does not scatter in the ice.  Since muons move faster than photons in the ice, they are likely to produce acausal photons.  The residual $t_{res}$ is calculated for each photon.
The variables are

\begin{description}
\item[PID variables] \hfill \\
\textbf{T1}: The smallest $t_{res}$ \\
\textbf{T2}: The number of hits in $\rm -200~ns< t_{res}<20~ns$ \\
\textbf{T3}: The distance that the cascade vertex moves when it is reconstructed, after omitting the acausal hits from the reconstruction.  \\
\textbf{T4}: $RLLH_{vertex}$ \\
\end{description}

The PID-BDT output shown in Fig.~\ref{fitpid} agrees 
well with the simulation expectation and shows good separation 
between $\nu_\mu$~CC and $\nu_e$-like events. 
Monte Carlo studies show that the $\nu_\mu$~CC identification improves 
at higher energies as the muon track becomes more visible. 
\begin{figure}[!t]
  \centering
  \includegraphics[width=3.5in]{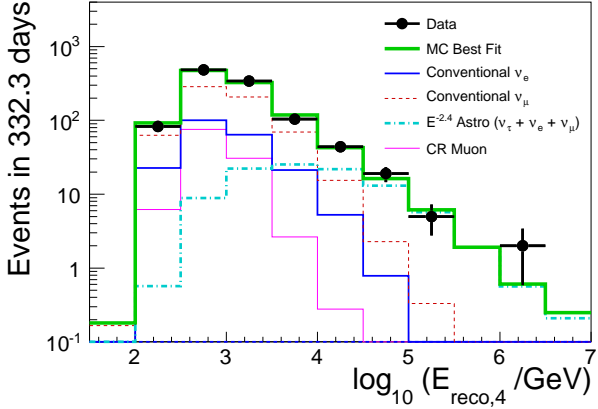}
  \caption{Reconstructed energy distribution for the baseline best fit. 
    Note that the prompt component is fit to zero. }
  \label{fite}
\end{figure}
\begin{figure}[!t]
  \centering
  \includegraphics[width=3.5in]{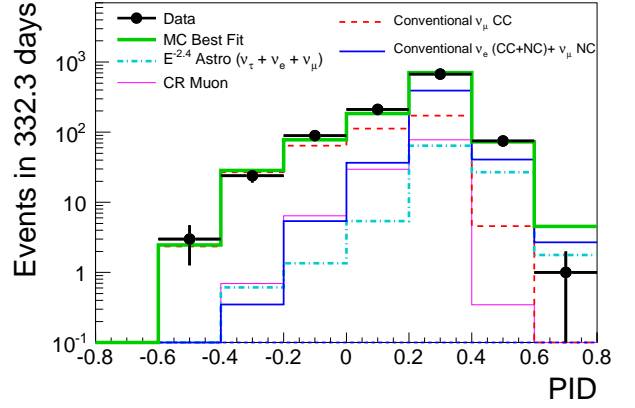}
  \caption{PID-BDT output distribution for the baseline best fit.
    The conventional $\nu_\mu$~CC as a hybrid component and 
    the conventional cascade component including $\nu_e$ and $\nu_\mu$~CC are plotted 
    separately. Events with high PID-BDT scores are cascades.
    The prompt component is fit to zero.
  }
  \label{fitpid}
\end{figure}
\begin{figure}[!t]
  \centering
  \includegraphics[width=3.5in]{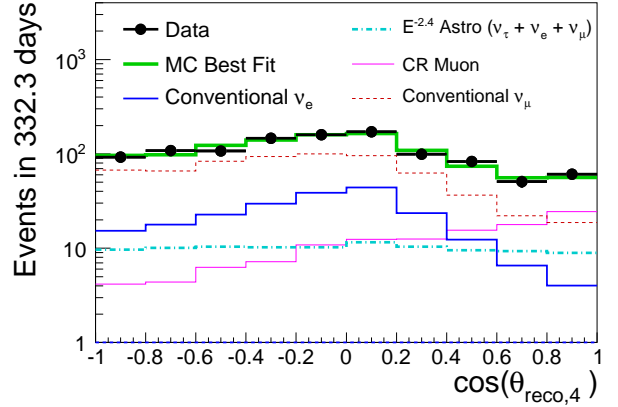}
  \caption{Reconstructed zenith angle for the baseline best fit.
    The prompt component is fit to zero.}
  \label{fitzen}
\end{figure}

\section{VI. Flux Measurement Method}

For measuring the atmospheric neutrino fluxes, the data was histogrammed in three dimensions $-$ energy, zenith angle and PID observable.
Three different fits were performed to the final data sample, to test different physics parameters. The first (baseline fit) was a straightforward measurement of the $\nu_e$ spectrum, assuming the spectral shape of the model. The second subdivided the energy spectrum, to make a binned measurement of the $\nu_e$ flux versus the neutrino energy. The third fit was similar to the baseline fit, but allowed the kaon to pion ratio to vary. These fits include the systematic uncertainties using a profile likelihood approach.

The parameters of the baseline fit are shown in Table~\ref{table_fit}.
\begin{table}[t]
\begin{center}
\caption{Baseline fit result table. The best fit includes statistical and systematic uncertainties 
  at 68\% C.L.}
\begin{tabular}{c|c}
\hline\hline
Parameters         & Best fit  \\
\hline
CR muon            &   $115^{+28}_{-27}$  events                      \\
Conventional $\nu_\mu$    &   $1.0^{+0.2}_{-0.1}$ $\times$ modified Honda   \\
Conventional $\nu_e$      &   $1.3^{+0.4}_{-0.3}$ $\times$ modified Honda  \\
Prompt Normalization            &   $0.0^{+3.0}_{-0.0}$ $\times$ modified ERS   \\
Astrophysical flux    &   $3.2^{+1.1}_{-0.9}\times10^{-18}$ \\
   &   $\times \rm GeV^{-1} cm^{-2} sr^{-1} s^{-1} (E_\nu / 10^5 GeV)^{-\gamma}$  \\
Astrophysical $\gamma$    &   $2.4^{+0.1}_{-0.2}$                       \\
\hline
Optical Efficiency       &   $-$2.2\%                                 ($\pm$10\% prior)\\
Ice Parameters     &   $+1.0~\sigma$ at scattering +10\%\\
\hline\hline
\end{tabular}
\label{table_fit}
\end{center}
  \vspace{-5mm}
\end{table}
Six physics parameters are used: conventional $\nu_\mu$ 
and $\nu_e$ normalizations relative to the modified Honda flux, a CR muon normalization, 
a total prompt ($\nu_\mu + \nu_e$) normalization with respect to the modified 
ERS flux, an astrophysical normalization ($\phi_0$) and an astrophysical spectral index ($\gamma$).  Figures 6 to 8 show one-dimensional projections 
of these histograms with the bin numbers and their ranges used in the fitter, along with the baseline fit results.

In the second fit, the conventional $\nu_e$ flux component is 
further divided into four smaller energy ranges spanning 100~GeV to 100~TeV, introducing three additional physics parameters.  Because the region above 100~TeV is dominated by the astrophysical component, it is not used in this separate fit.

Finally, in the third fit, a kaon fraction parameter and 
a total conventional ($\nu_\mu + \nu_e$) normalization are introduced
in order to remove any correlation between conventional $\nu_\mu$ and $\nu_e$.

For these fits, the likelihood $L$ is constructed with a Poissonian component 
for the physics parameters and Gaussian components for the systematic parameters. 
Best fit results are obtained by minimizing the negative logarithm of $L$,
\begin{equation} \label{eq2}
  - 2\ln L=2\sum_{\bf{k}} \big(\mu_{\bf{k}}-n_{\bf{k}}\ln\mu_{\bf{k}}\big) + 
  \sum_{m}\Big(\frac{l_m-\hat{l}_m}{\sigma_{l_m}}\Big)^2
\end{equation}
All physics parameters are unconstrained in the fitting process 
while the two systematic parameters are restricted by the priors that quantify their estimated precision.
The total expected count $\mu$ is a sum of each component's contribution which depends on the fit parameters. The index $\bf{k}$ iterates 
over the histogram bins, and the number of observed events in bin $\bf{k}$ is $n_{\bf{k}}$. 
The expected count also depends on given systematic parameters. 
The two systematic parameters ($m$) have a central value ($\hat{l}$) and an uncertainty ($\sigma_l$). 
The 68\% parameter uncertainties are determined by scanning $-2\ln L$
up to one unit from the best fit likelihood.

\section{VII. Systematic Uncertainties}
Systematics uncertainties arise due 
to imperfect modelling of our detector which can affect analysis results.
The two most important detector systematics are included in the fitter. They are the total optical efficiency of a DOM 
and the optical properties of surrounding ice (scattering and absorption lengths). 
The sizes of these systematic errors are estimated from laboratory measurements 
of DOMs and ice measurements with \textit{in-situ} devices. 
A few simulations with different input assumptions on the systematic effects 
are performed and their event rates and shapes are compared with 
those in the nominal Monte Carlo at the final analysis level.

Simulations with modified optical DOM efficiency result 
in a different event rate globally but show little 
change in the shape of the analysis observable distributions. 
Since the normalization of an assumed physics model 
translates directly to a flux of that model, this systematic 
uncertainty loosens the constraint on the flux. 
The impact on the event rate relative to the 
nominal value is parameterized using five simulations with 
the input efficiencies ranging from $-10\%$ to $+10\%$. 
This results in output event rate changes in the range $[-20\%,+10\%]$, 
matching the high event rate for the high efficiency input (see Fig.~\ref{syst}). 
\begin{figure}[!t]
  \centering
  \includegraphics[width=3.in]{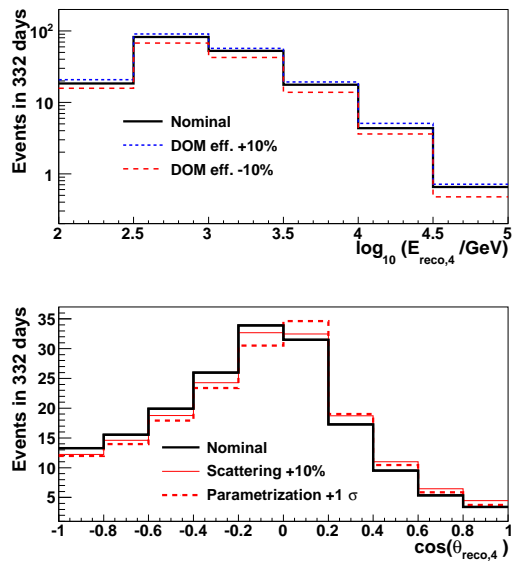}
  \caption{Effects of systematic variations in final 
    level $\nu_e$ sample. The upper panel shows the DOM efficiency impact 
    on the reconstructed energy compared to the nominal prediction. 
    The lower panel shows the ice systematic impact 
    due to an increased scattering on the reconstructed zenith angle.}
  \label{syst}
\end{figure}
The asymmetric change in event rate with varying optical efficiency is a combined effect of the changing number of observed photoelectrons, coupled with the analysis selection of higher quality events. 

The ice systematics alter the shape of the zenith 
angle distribution. A global increase in light scattering 
tilts the zenith angle reconstruction 
downwards (see Fig.~\ref{syst}). 
This is a consequence of losing non-scattered hits that 
are crucial in the cascade direction reconstruction.
A 10\% increase in scattering coefficient
degrades the angular resolution by about 2 degrees.
However, the change in the absorption coefficients 
has a smaller impact on the zenith angle shape. 
The change in zenith angle distribution is modelled as a parameterization of the ice model by reweighting the event rate.
This model changes the nominal event rate by $-10\%$ 
in the upward going direction and by $+10\%$ in the downward 
going region for a positive one-sigma shift in the fitter ($\frac{l-\hat{l}}{\sigma}=1$) (see Fig.~\ref{syst}).

We have investigated other systematic effects arising from the neutrino-nucleon 
cross-section and cosmic ray spectral slope. 
The theoretical uncertainties from the neutrino-nucleon deep inelastic 
cross-section~\cite{CooperSarkar:2011pa,Connolly:2011vc} are relatively small compared to the other
systematic uncertainties. We assume a 3\% cross-section uncertainty, following Ref. \cite{Aartsen:2013vca}.
The systematic impact of the cross-section acts as a simple normalization 
in the energy region of this analysis and is strongly correlated 
with the DOM efficiency parameterization. 
The cosmic ray spectral slope has a small impact, compared to the detector-related systematic uncertainties. 
Additionally, the systematic effect on the zenith angle shape due to a seasonal temperature variation and the atmospheric self veto calculation are similar to ice systematics and absorbed by the parameterization of ice systematics.

\section{VIII. Results and Discussion}
A total of 1078 events are observed after unblinding the full data set.
The cascade candidates are distributed evenly throughout the year and  
no events are coincident in time with the IceTop triggers.
Figure~6 shows the energy spectrum.  The average reconstructed energy is $\langle E\rangle\sim$1.7~TeV 
with 970 events (90\%) between 278~GeV and 13.5~TeV. Above 10~TeV, 70 events are detected.
Of the total, 57\% are reconstructed as upward-going.
The baseline fit results are shown in Table~\ref{table_fit} with the total uncertainties.
The 1-dimensional projected distributions for the best fit  zenith angle and PID variables 
are shown in Figures 7 and 8.

Figure~\ref{fite} shows the energy spectra of the different components.
The 115$^{+28}_{-27}$ CR muon events (11\% of the total)
follow the distribution estimated from the background-dominated data region.
At the lowest energies, the background contamination is track-like, 
and mostly downward-going. 
These events are not easily detectable by visual inspection
and are most likely stealth muon events with little veto information.

The conventional $\nu_\mu$ and $\nu_e$ components are mainly horizontal, at
low energies. 
The energy and zenith angle distributions are similar, 
so much of the $\nu_e/\nu_\mu$ separation power comes from the PID observable. 
The PID separates $\nu_\mu$~CC from other events which have no trace of a muon track : NC events and $\nu_e$~CC.
The $\nu_e$~CC and $\nu_\mu$~NC events are indistinguishable. 
The fit finds the $\nu_\mu$ normalization at $1.0^{+0.2}_{-0.1}$ $\times$ modified Honda (645 events)
and the $\nu_e$ normalization at $1.3^{+0.4}_{-0.3}$ $\times$ modified Honda (215 events).
The flux ratio of $\nu_\mu$ to $\nu_e$ at 1.7~TeV 
is  $16.9^{+6.4}_{-4.0}$ compared with the Honda prediction of 23 and 
 the Bartol~\cite{Bartol-2004} prediction of 14. The models use different assumptions 
about the primary cosmic-ray spectrum and the treatment of 
kaons~\cite{IntHonda-2007,BartolError-2006}.

The $\nu_e$ to $\nu_\mu$ ratio depends on the kaon to pion ratio in cosmic-ray air showers.
One of the major uncertainties in the $K:\pi$ ratio is due to associated production via reactions like $p+N \rightarrow \Lambda + K^+$.
A higher rate of associated production leads to fewer $\overline\nu_e$ and more $\nu_e$ at energies 
above 1~TeV~\cite{Gaisser:2002jj}. Since the $\overline\nu_e$ and $\nu_e$ have different interaction cross-sections in the ice, this will lead to a smaller amount in the total $\nu_e$ rate, resulting in higher $\nu_\mu/\nu_e$ ratio.
Both calculations suffer from large uncertainties regarding kaon production at these energies.

The statistical uncertainties on the $\nu_\mu$ and $\nu_e$ normalizations 
are estimated to be 8.6\% and 20\%, respectively as determined by running 
the fitter without the systematic parameters included.
The conventional normalization results are consistent with Honda 
predictions and the significance contours of the conventional normalization fit are shown in Fig.~\ref{contour_conv}. 
\begin{figure}[!t]
  \centering
  \includegraphics[width=3.0in]{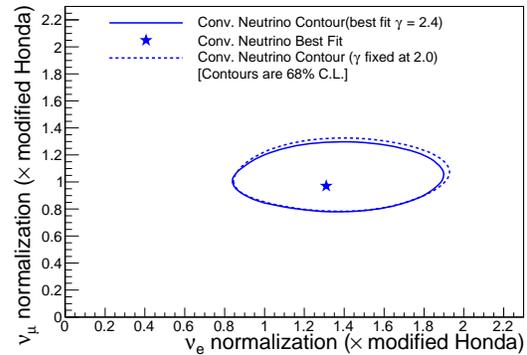}
  \caption{Best fit contours for the conventional flux at 68\% C.L. 
    The baseline fit with the astrophysical spectral index $\gamma$ free is shown with solid 
    blue line. An alternative fit with the index 
    fixed at $\gamma=2.0$ is shown as a dotted line. 
  }
  \label{contour_conv}
\end{figure}
Overall, the CR muons and the conventional neutrinos are not correlated with prompt
or astrophysical components.
As can be seen in Fig.~\ref{contour_conv}, the change in 
conventional normalization with the astrophysical model is minimal.

On the other hand, the prompt normalization is strongly 
influenced by astrophysical models.
The fit for the prompt normalization is zero 
with the 68\% confidence upper limit at 3.0$\times$modified ERS.
The best fit astrophysical flux per flavor is 
$3.2^{+1.1}_{-0.9}\times 10^{-18}$ $\rm GeV^{-1} cm^{-2} sr^{-1} s^{-1} (E_\nu / 10^5 GeV)^{-\gamma}$ 
with $\gamma$ = 2.4$^{+0.1}_{-0.2}$.
The relationship between the fits for the prompt flux and astrophysical 
models is shown in Fig.~\ref{limit}.
\begin{figure}[!t]
  \centering
  \includegraphics[width=3.5in]{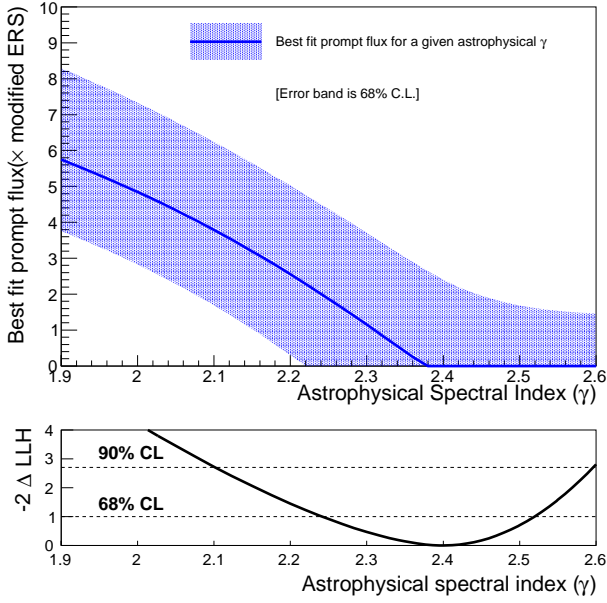}
  \caption{For a given astrophysical spectral index (x axis) 
    in the upper panel, the best fit prompt flux (blue line) and 
    its errors (band at 68\% C.L.) from the profile likelihood scan are obtained. 
    The bottom panel shows the range of allowed region of the index parameter from the full fit.}
  \label{limit}
\end{figure}
As the astrophysical spectral index softens, the shapes of the prompt
and astrophysical components in the observable space become similar.  In the limit of identical indices, the main way to separate these two components is via self-vetoing; down-going prompt neutrinos will be accompanied by muons which will cause the event to be rejected.  This will show up as a change in the zenith angle distribution, with down-going events suppressed, in contrast to the astrophysical component, which will remain isotropic. 

The presence of very high energy events ($\sim$1~PeV) in the downward region favors the astrophysical component over the prompt component.  It should be noted that the presence of the cosmic-ray knee introduces a kink into the prompt component spectrum.  As Fig. \ref{nuflux} shows, at energies above a few hundred TeV, this kink further reduces the prompt component.

Since the fit results for the conventional components are not influenced by 
the prompt or astrophysical components, 
we obtain the conventional $\nu_e$ spectrum independent of assumptions about the other components. 
A separate fit is performed by introducing conventional $\nu_e$ components 
divided into four true energy ranges
while keeping all of the other components unchanged.
The resulting best-fit normalizations in each range produce
the neutrino fluxes as shown in Fig.~\ref{nuflux} and Table~\ref{fluxtable}. 
The fit finds good agreement with models of the conventional $\nu_e$ flux.
The other components in the fit show consistent values when compared to 
the previous baseline fit.
\begin{figure}[!t]
  \vspace{5mm}
  \centering
  \includegraphics[width=3.5in]{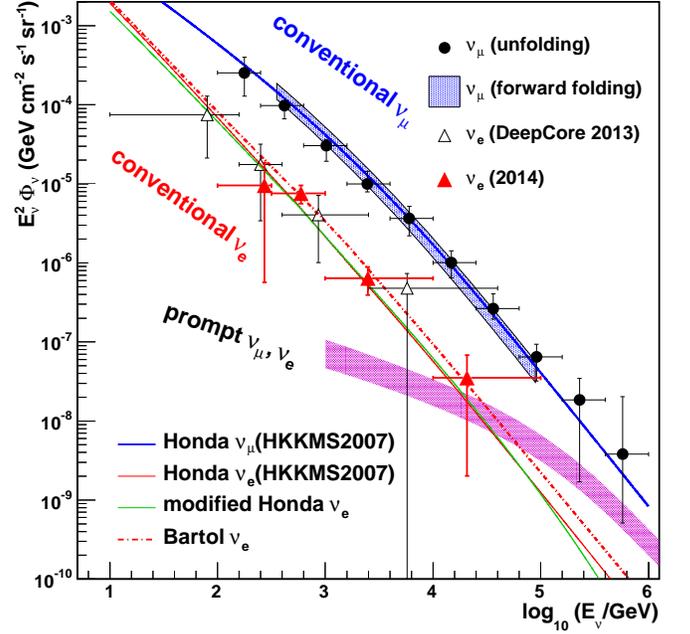}
  \caption{The atmospheric $\nu_e$ flux result (shown as red filled triangles). 
    Markers indicate the IceCube measurements of the atmospheric 
    neutrino flux while lines show the theoretical models. 
    The black circles and the blue band come from the through-going 
    upward $\nu_\mu$ analyses~\cite{Aartsen:2014qna,Diffuse-Sean-2011}. 
    The open triangles show the $\nu_e$ measurement with the IceCube-DeepCore dataset~\cite{Aartsen:2012uu}.
    The magenta band shows the modified ERS prediction.}
  \label{nuflux}
\end{figure}

\begin{table}[t]
  \vspace{-5mm}
  \begin{center}
    \caption{The results of the binned (`second') fit to the $\nu_e$ flux for an $E^{-2}$ spectrum, in four energy bins.}
    \begin{tabular}{c|c|cc}
      \hline
      $\rm \log_{10}E_\nu^{min}\!-\!\log_{10}E_\nu^{max} $& $\rm \langle E_{\nu} \rangle$(GeV) &$\rm E^2_\nu\Phi_\nu$($\rm GeV~cm^{-2}s^{-1}sr^{-1}$)& \\
\hline
2.0 $-$ 2.5   & 270   & $(1.0\pm 0.9) \times 10^{-5}$     &      \\
2.5 $-$ 3.0   & 590   & $(7.6\pm 1.9) \times 10^{-6}$     &      \\
3.0 $-$ 4.0   & $2.5\times 10^{3}$  & $(6.4\pm 2.6) \times 10^{-7}$     &      \\
4.0 $-$ 5.0   & $20.7\times 10^{3}$ & $(3.5\pm 3.3) \times 10^{-8}$     &      \\
\hline
\end{tabular}
\label{fluxtable}
\end{center}
  \vspace{-5mm}
\end{table}

The relatively high conventional $\nu_e$ flux normalization measured in the first fit 
can be further examined by varying the relative contribution 
from $\pi$ and $K$ to the conventional neutrino fluxes.
In a third fit, we introduce an extra fit parameter ($\xi$)
which modifies the $K$ contributions in Eq.~\ref{eq7}
and in Eq.~\ref{eq8} simultaneously.
\begin{eqnarray}  \label{eq7}
  \Phi_{\nu_\mu}(\xi) &=& C \cdot E^{-2.65}_{\nu_\mu} \cdot ( w_{\pi} + \xi \cdot w_{K}  )
\end{eqnarray}
\begin{eqnarray}  \label{eq8}
 \Phi_{\nu_e}(\xi) &=& C'\cdot E^{-2.65}_{\nu_e} \cdot \xi \cdot w_{K'}
\end{eqnarray}

A value of $\xi=1$ corresponds to the standard expectations 
based on the modified Honda model and a value of $\xi>1$ corresponds to increased kaon production.
As the conventional $\nu_\mu$ and $\nu_e$ flux normalizations are fixed to the baseline model, 
$\xi$ probes the deviations from the model due to relative $K$ contribution.
The $\nu_e$ normalization $C'$ and the kaon weight $w_{K'}$ are fixed at the Honda flux.
For the $\nu_\mu$ part, while the change in $\xi$ corresponds to a change in shape of the energy distribution,
the total number of $\nu_\mu$ events is fixed to the baseline expectation due to the change in $\xi$.
On the other hand, an increase in the $K$ contribution to $\nu_e$ causes the number of events
in the $\nu_e$ prediction to increase while the shape is unchanged.
This is because $\nu_e$ comes mostly from $K$ in these energies.
The $\nu_e$ flux from $\pi\rightarrow\mu\rightarrow\nu_e$ decays 
is negligible, so there is little shape change in the $\nu_e$ energy spectrum due to $\pi$.
This fit finds $\xi = 1.3^{+0.5}_{-0.4}$ with respect to the modified Honda flux.

The central value of the $K$ content is above standard calculations, although the errors
are large.  Current models of cosmic ray interactions may underestimate the strange quark content in the air shower.  Enhanced strangeness production has been measured in nuclear 
collisions at Relativistic Heavy Ion Collider~\cite{Klein:2006pp}, and air shower 
experiments also measure higher muon contents for inclined showers compared 
with the predictions from existing hadronic interaction 
models~\cite{ThePierreAuger:2013mga,Aab:2014pza,ThePierreAuger:2013eja}. 

\section{IX. Conclusions}

In conclusion, we obtained a sample of 1078 cascade events in the analysis of one year of data from the completed IceCube detector.  This sample is used to measure the conventional atmospheric $\nu_e$ flux.  The analysis is designed so that the conventional neutrino result is 
largely unaffected by the prompt neutrino flux and/or the astrophysical models. 
The analysis extends previous measurements ~\cite{Aartsen:2012uu} of the $\nu_e$ flux to higher energies, and provides higher precision. The first analysis with only the DeepCore region as a fiducial volume was optimized in obtaining a large number of lower energy events. Therefore, the improvement comes from a better event selection by expanding the fiducial volume for higher energy events and a three dimensional likelihood method including particle identification at higher energies. 

The conventional $\nu_e$ spectrum was measured
between 0.1~TeV and 100~TeV. The measured $\nu_e$ flux was $1.3^{+0.4}_{-0.3} \times$ modified Honda prediction which includes a model of the cosmic-ray knee and a correction to account for self-vetoing, whereby an atmospheric neutrino is accompanied by muons from the same shower, causing it to fail the event selection.  An unfolding was used to determine the $\nu_e$ flux  in four energy bins. 

In addition to the conventional $\nu_e$ spectrum measurements, we find that the result for the prompt 
component strongly depends on the assumed astrophysical models. 
The analysis fits the prompt flux at $0.00^{+3.0}_{-0.0}\times$modified ERS, 
together with the astrophysical flux per flavor at 
$3.2^{+1.1}_{-0.9} \times 10^{-18}$ $\rm GeV^{-1} cm^{-2} sr^{-1} s^{-1} (E_\nu / 10^5 GeV)^{-\gamma}$ 
with $\gamma = 2.4^{+0.1}_{-0.2}$ at 68\% C.L. 
The uniqueness of the prompt compared to soft astrophysical components is two-fold: 
a shape difference in energy due to the presence of cosmic ray 
knee and a shape difference in zenith angle
due to the impact of the self veto. 

The analysis also finds a slightly higher $K$ contribution than in current models, at $1.3^{+0.5}_{-0.4} \times$ modified Honda. 
The measured neutrino flux ratio $\nu_\mu/\nu_e = 16.9^{+6.4}_{-4.0}$ 
at the mean neutrino energy of 1.7~TeV, is below the prediction of the Honda model, but slightly above the prediction of the Bartol model. 

At energies above a few TeV, additional data, as would be provided by a multi-year analysis,
would allow for a more precise measurement.

\begin{acknowledgments}
  \section{X. acknowledgments}
We acknowledge the support from the following agencies:
U.S. National Science Foundation-Office of Polar Programs,
U.S. National Science Foundation-Physics Division,
University of Wisconsin Alumni Research Foundation,
the Grid Laboratory Of Wisconsin (GLOW) grid infrastructure at the University of Wisconsin - Madison, the Open Science Grid (OSG) grid infrastructure;
U.S. Department of Energy, and National Energy Research Scientific Computing Center,
the Louisiana Optical Network Initiative (LONI) grid computing resources;
Natural Sciences and Engineering Research Council of Canada,
WestGrid and Compute/Calcul Canada;
Swedish Research Council,
Swedish Polar Research Secretariat,
Swedish National Infrastructure for Computing (SNIC),
and Knut and Alice Wallenberg Foundation, Sweden;
German Ministry for Education and Research (BMBF),
Deutsche Forschungsgemeinschaft (DFG),
Helmholtz Alliance for Astroparticle Physics (HAP),
Research Department of Plasmas with Complex Interactions (Bochum), Germany;
Fund for Scientific Research (FNRS-FWO),
FWO Odysseus programme,
Flanders Institute to encourage scientific and technological research in industry (IWT),
Belgian Federal Science Policy Office (Belspo);
University of Oxford, United Kingdom;
Marsden Fund, New Zealand;
Australian Research Council;
Japan Society for Promotion of Science (JSPS);
the Swiss National Science Foundation (SNSF), Switzerland;
National Research Foundation of Korea (NRF);
Danish National Research Foundation, Denmark (DNRF)
\end{acknowledgments}

\end{document}